\documentclass[a4paper]{article}
\usepackage{graphicx}
\usepackage{amsmath}

\begin{document}

\title{Simultaneous amplification and non-symmetric amplitude damping of two-mode
Gaussian state}
\author{Xiao-yu Chen,Li-zhen Jiang, Ji-wu Chen \\
{\small {Lab. of Quantum Information, China Institute of Metrology,
Hangzhou, 310034, China}}}
\date{}
\maketitle

\begin{abstract}
The evolution of two-mode Gaussian state under symmetric amplification,
non-symmetric damping and thermal noise is studied. The time dependent
solution of the state characteristic function is obtained. The separability
criterions are given for the final state of weak amplification as well as
strong amplification.


\end{abstract}

\section{Introduction}

In all practical instances the information and entanglement contained in a
given quantum state of the system, so precious for the realization of any
specific task, is constantly threatened by the unavoidable interaction with
the environment. Such an interaction entangles the system of interest with
the environment, causing any amount of information to be scattered and lost
in the environment. The overall process, corresponding to a non unitary
evolution of the system, is commonly referred to as decoherence. To overcome
the loss, parameter amplifier is added to the system. We in this paper will
treat the simultaneous actions of amplitude damping and parameter
amplification to two-mode Gaussian state.

The density matrix obeys the following master equation \cite{Walls} $\frac{%
d\rho }{dt}=-\frac i\hbar [H,\rho ]+\mathcal{L}\rho ,$with the quadratic
Hamiltonian $H=\hbar \sum_{jk}\frac i2(\eta _{jk}a_j^{\dagger }a_k^{\dagger
}-\eta _{jk}^{*}a_ja_k)$, where $\eta $ is a complex symmetric matrix
(parameter amplifier matrix). $\mathcal{L}_1\rho =\sum_j\frac{\Gamma _j}2\{(%
\overline{n}_j+1)L[a_j]\rho +\overline{n}_jL[a_j^{\dagger }]\rho \},$where
the Lindblad super-operators are defined as $L[\widehat{o}]\rho \equiv $ $2%
\widehat{o}\rho \widehat{o}^{\dagger }-\widehat{o}^{\dagger }\widehat{o}\rho
-\rho \widehat{o}^{\dagger }\widehat{o}$ $,\Gamma _j$ is the amplitude
damping coefficient of $jth$ mode, $\overline{n}_j$ is the average thermal
photon number of the environment. Any quantum state can be equivalently
specified by its characteristic function. Every operator $\mathcal{A}\in
\mathcal{B(H)}$ is completely determined by its characteristic function $%
\chi _{\mathcal{A}}:=tr[\mathcal{AD}(\mu )]$ \cite{Petz}, where $\mathcal{D}%
(\mu )=\exp (\mu a^{\dagger }-\mu ^{*}a)$ is the displacement operator, with
$\mu =[\mu _1,\mu _2,\cdots ,\mu _s]$ $,a=[a_1,a_2,\cdots ,a_s]^T$ and the
total number of modes is $s.$ It follows that $\mathcal{A}$ may be written
in terms of $\chi _{\mathcal{A}}$ as \cite{Perelomov}: $\mathcal{A}=\int
[\prod_i\frac{d^2\mu _i}\pi ]\chi _{\mathcal{A}}(\mu )\mathcal{D}(-\mu ).$
The density matrix $\rho $ can be expressed with its characteristic function
$\chi $. $\chi =tr[\rho \mathcal{D}(\mu )]$ . The master equation can be
transformed to the diffusion equation of the characteristic function, it is
\cite{Chen1} \cite{Chen}
\begin{equation}
\frac{\partial \chi }{\partial t}=-\sum_{jk}(\eta _{jk}\mu _j^{*}\frac{%
\partial \chi }{\partial \mu _k}+\eta _{jk}^{*}\mu _j\frac{\partial \chi }{%
\partial \mu _k^{*}})-\frac 12\sum_j\Gamma _j\{\left| \mu _j\right| \frac{%
\partial \chi }{\partial \left| \mu _j\right| }+(2\overline{n}_j+1)\left|
\mu _j\right| ^2\chi \}.  \label{qw1}
\end{equation}

\section{ The parametric amplifier and the amplitude damping}

The solution of the diffusion equation of the characteristic function can be
completely worked out for Gaussian state in the case of real parameter
amplifier matrix $\eta $. We will consider real $\eta $ in the following. If
the initial state is Gaussian, its characteristic function has the form of $%
\chi (\mu ,\mu ^{*},0)=\exp [\mu m^{\dagger }(0)-\mu ^{*}m^T(0)-\frac 12(\mu
,-\mu ^{*})\gamma (0)(\mu ^{*},-\mu )^T],$ the state will keep to be a
Gaussian state in later evolution, where $m$ is the first moment and is
irrelevant to entanglement, $\gamma $ is the complex correlation matrix
(CM). The time evolution of the complex CM for real amplifier matrix $\eta $
is \cite{Chen}
\begin{equation}
\gamma (t)=\left[
\begin{array}{ll}
M & -N \\
-N & M
\end{array}
\right] (\gamma (0)-\left[
\begin{array}{ll}
\alpha & \beta ^{*} \\
\beta & \alpha ^{*}
\end{array}
\right] )\left[
\begin{array}{ll}
M & -N \\
-N & M
\end{array}
\right] +\left[
\begin{array}{ll}
\alpha & \beta ^{*} \\
\beta & \alpha ^{*}
\end{array}
\right] .  \label{qw2}
\end{equation}
where $M$ and $N$ are the solutions of the following matrix equations $\frac{%
dM}{dt}=-\eta N-\frac \Gamma 2M,\frac{dN}{dt}=-\eta M-\frac \Gamma 2N,$with $%
\Gamma =diag\{\Gamma _1,\Gamma _2,\cdots ,\Gamma _s\}.$ The solution is $%
M=\frac 12[\exp (-\eta t-\frac{\Gamma t}2)+\exp (\eta t-\frac{\Gamma t}2)],$
$N=\frac 12[\exp (-\eta t-\frac{\Gamma t}2)-\exp (\eta t-\frac{\Gamma t}2)].$
The constant matrices $\alpha $ and $\beta $ in \ref{qw2} have the behaviors
$\alpha ^{\dagger }=\alpha ,$ $\beta =\beta ^T$, they are the solutions of
the following matrix equations
\begin{eqnarray}
2(\eta \alpha +\alpha ^{*}\eta )-\Gamma \beta -\beta \Gamma &=&0,
\label{we3} \\
\Gamma \alpha +\alpha \Gamma -2\eta \beta -2\beta ^{*}\eta -\Gamma (%
\overline{n}+\frac 12)-(\overline{n}+\frac 12)\Gamma &=&0.  \label{we4}
\end{eqnarray}
where $\overline{n}=diag\{\overline{n}_1,\overline{n}_2,\cdots ,\overline{n}%
_s\}.$ The one mode solution has been known for a long time (see \cite
{Corney} and references therein).

For the two-mode situation, the real amplifier matrix $\eta =\eta _0\sigma
_0+\eta _1\sigma _1+\eta _3\sigma _3,$where $\sigma _0=I_2,\sigma _1,\sigma
_3$ are Pauli matrices. $M$ and $N$ can be simplified to
\begin{eqnarray}
M &=&\frac 12e^{-C_1t}[\cosh (B_1t)\sigma _0-\sinh (B_1t)\overrightarrow{%
\sigma }\cdot \overrightarrow{b_1}]+\frac 12e^{-C_2t}[\cosh (B_2t)\sigma
_0-\sinh (B_2t)\overrightarrow{\sigma }\cdot \overrightarrow{b_2}],
\label{qw3} \\
N &=&\frac 12e^{-C_1t}[\cosh (B_1t)\sigma _0-\sinh (B_1t)\overrightarrow{%
\sigma }\cdot \overrightarrow{b_1}]-\frac 12e^{-C_2t}[\cosh (B_2t)\sigma
_0-\sinh (B_2t)\overrightarrow{\sigma }\cdot \overrightarrow{b_2}].
\label{qw4}
\end{eqnarray}
where $C_{1,2}=\pm \eta _0+\frac 14(\Gamma _1+\Gamma _2);$ $B_{1,2}=\sqrt{%
\eta _1^2+(\frac 14(\Gamma _1-\Gamma _2)\pm \eta _3)^2};$ $\overrightarrow{b}%
_{1,2}=(\pm \eta _1,0,\pm \eta _3+\frac 14(\Gamma _1-\Gamma _2))/B_{1,2}.$
We will consider the case of symmetric noise, that is, $\overline{n}=%
\overline{n}_0\mathbf{I}_{2.}$The solutions of equations (\ref{we3}) and (%
\ref{we4}) are given in the appendix.

\section{ The inter-mode amplifier}

The algebra equation of $\alpha $ and $\beta $ in two mode system is
complicated in general situation. To investigate the entanglement property
of the amplifier, we will first consider the case of $\eta _0=\eta _3=0$
which corresponds to inter-mode amplification alone. Thus $\eta =\eta
_1\sigma $. The solution is (see Appendix)
\begin{eqnarray}
\alpha &=&\frac{\overline{n}_0^{\prime }}{1-\Gamma _3^{\prime 2}-\eta
_1^{\prime 2}}[(1-\Gamma _3^{\prime 2})\sigma _0+\Gamma _3^{\prime }\eta
_1^{\prime 2}\sigma _3], \\
\beta &=&\frac{\overline{n}_0^{\prime }\eta _1^{\prime }(1-\Gamma _3^{\prime
2})}{1-\Gamma _3^{\prime 2}-\eta _1^{\prime 2}}\sigma _1,
\end{eqnarray}
where $\overline{n}_0^{\prime }=\overline{n}_0+\frac 12,$ $\Gamma
_3^{^{\prime }}=\Gamma _3/\Gamma _0,\eta _1^{\prime }=2\eta _1/\Gamma _0.$
In the assumption of $\eta _0=\eta _3=0,$ we have $C_1=C_2=\Gamma _0/2;$ $%
B_1=B_2=\sqrt{\left( \Gamma _3/2\right) ^2+\eta _1^2}=k\Gamma _0/2$, with $k=%
\sqrt{\Gamma _3^{^{\prime }2}+\eta _1^{\prime 2}}.$ Denote $t^{\prime
}=\Gamma _0t/2,$hence
\begin{eqnarray}
M &=&e^{-t^{\prime }}[\cosh (kt^{\prime })\sigma _0-\sinh (kt^{\prime })%
\frac{\Gamma _3^{\prime }}k\sigma _3] \\
N &=&-e^{-t^{\prime }}\sinh (kt^{\prime })\frac{\eta _1^{\prime }}k\sigma _1
\end{eqnarray}

For the case of weak amplifier, $k<1$, that is , $\eta _1^2<\frac 14(\Gamma
_0^2-\Gamma _3^2),$ when $t\rightarrow \infty ,$ we have $M,N\rightarrow 0.$
The state will tend to a Gaussian state which is characterized by the
residue complex CM $\gamma (\infty )$ $=\left[
\begin{array}{ll}
\alpha  & \beta ^{*} \\
\beta  & \alpha ^{*}
\end{array}
\right] $. The Peres-Horodecki criterion for separability \cite{Simon} \cite
{Duan} will be \cite{Chen}
\begin{equation}
\det \gamma _a\det \gamma _b+(\frac 14-\left| \det \gamma _c\right|
)^2-tr(\gamma _a\sigma _3\gamma _c\sigma _3\gamma _b\sigma _3\gamma
_c^{\dagger }\sigma _3)\geq \frac 14(\det \gamma _a+\det \gamma _b),
\end{equation}
where
\begin{equation}
\alpha =\left[
\begin{array}{ll}
\alpha _a & \alpha _c \\
\alpha _c^{*} & \alpha _b
\end{array}
\right] ,\text{ }\beta =\left[
\begin{array}{ll}
\beta _a & \beta _c \\
\beta _c & \beta _b
\end{array}
\right] ,\gamma _i^{\prime }=\left[
\begin{array}{ll}
\alpha _i & \beta _i^{*} \\
\beta _i & \alpha _i^{*}
\end{array}
\right] ,\text{ }i=a,b,c.
\end{equation}
Then $\alpha _a=$ $\frac{\overline{n}_0^{\prime }}{1-k^2}(1-\Gamma
_3^{\prime 2}+\Gamma _3^{\prime }\eta _1^{\prime 2}),$ $\alpha _b=\frac{%
\overline{n}_0^{\prime }}{1-k^2}(1-\Gamma _3^{\prime 2}-\Gamma _3^{\prime
}\eta _1^{\prime 2}),$ $\alpha _c=0;$ $\beta _c=\frac{\overline{n}_0^{\prime
}}{1-k^2}\eta _1^{\prime }\left( 1-\Gamma _3^{\prime 2}\right) ,$ $\beta
_a=\beta _b=0.$ The state is a x-p symmetric Gaussian state\cite{Jiang}
whose Gaussian relative entropy of entanglement can be obtained \cite{Chen2}%
. The separability criterion now is $\alpha _a^2\alpha _b^2+(\frac 14-\beta
_c^2)^2-2\alpha _a\alpha _b\beta _c^2\geq \frac 14(\alpha _a^2+\alpha _b^2),$
which can be reduced to $(\alpha _a-\frac 12)(\alpha _b-\frac 12)-\beta
_c^2\geq 0,$ that is
\begin{equation}
\lbrack 1-\Gamma _3^{\prime 2}(2\overline{n}_0+1)^2]\eta _1^{\prime 2}\leq 4%
\overline{n}_0^2(1-\Gamma _3^{\prime 2}).  \label{qw5}
\end{equation}

For the case of strong amplifier, $k>1$, that is , $\eta _1^2>\frac
14(\Gamma _0^2-\Gamma _3^2),$ suppose the complex CM is $\gamma (t)$ $%
=\left[
\begin{array}{ll}
\alpha ^{\prime } & \beta ^{\prime *} \\
\beta ^{\prime } & \alpha ^{\prime *}
\end{array}
\right] $ at time $t,$ a direct calculation shows that $\alpha ^{\prime
}=diag\{\alpha _a^{\prime },\alpha _b^{\prime }\},$ $\beta ^{\prime }=\beta
_c^{\prime }\sigma _1$ with
\begin{eqnarray}
\alpha _a^{\prime } &=&\alpha _a+M_a^2(\frac 12-\alpha _a)+N_c^2(\frac
12-\alpha _b)+2M_aN_c\beta _c, \\
\alpha _b^{\prime } &=&\alpha _b+M_b^2(\frac 12-\alpha _b)+N_c^2(\frac
12-\alpha _a)+2M_bN_c\beta _c, \\
\beta _c^{\prime } &=&\beta _c-M_aN_c(\frac 12-\alpha _a)-M_bN_c(\frac
12-\alpha _b)-(M_aM_b+N_c^2)\beta _c,
\end{eqnarray}
where we have denoted $M=diag\{M_a,M_b\},$ $N=N_c\sigma _1;$ the
vacuum initial state is assumed. The state is still a x-p
symmetric Gaussian state. The separability criterion is $(\alpha
_a^{\prime }-\frac 12)(\alpha _b^{\prime }-\frac 12)-\beta
_c^{\prime 2}\geq 0,$ which can be written as
\begin{eqnarray}
&&\eta _1^{\prime 4}\{(K_1^2-1)(K_2^2-1)-(K_1K_2-1)^2(2\overline{n}%
_0+1)^2\Gamma _3^{\prime 2}\}  \nonumber \\
&&-\eta _1^{\prime 2}\{(K_1K_2-1)^2(1-\Gamma _3^{\prime 2})\Gamma _3^{\prime
2}(4\overline{n}_0^2-1)+4\overline{n}_0^2(K_1^2-1)(K_2^2-1)  \nonumber \\
&&-4\overline{n}_0\Gamma _3^{\prime 2}[K_2^2-\Gamma _3^{\prime
2}-2K_1K_2(1-\Gamma _3^{\prime 2})+K_1^2(1-K_2^2\Gamma _3^{\prime 2})]\}-4%
\overline{n}_0^2(K_1^2-1)(K_2^2-1)(1-\Gamma _3^{\prime 2})\Gamma _3^{\prime
2}  \nonumber \\
&\geq &0
\end{eqnarray}
where $K_1=e^{(k-1)t^{\prime }},K_2=e^{-(k+1)t^{\prime }}.$ When $t^{\prime
}\rightarrow \infty ,$ the separability criterion will be
\begin{equation}
\eta _1^{\prime 2}\leq 2\overline{n}_0(\overline{n}_0+\Gamma _3^{\prime 2}+%
\sqrt{\overline{n}_0^2+(2\overline{n}_0+1)\Gamma _3^{\prime 2}}).
\label{qw6}
\end{equation}
Inequalities (\ref{qw5}) (\ref{qw6}) are displayed in Fig.1 in a combined
form. The critical noise $\overline{n}_0$ is shown as a function of $\eta
_1^{\prime }$ and $\Gamma _3^{\prime }.$
\begin{figure}[tbp]
\includegraphics[ trim=0.000000in 0.000000in -0.138042in 0.000000in,
height=2.0081in, width=2.5097in ]{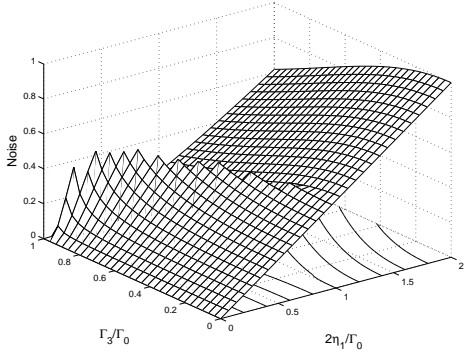}
\caption{The separability border of the amplification and non-symmetric
damping Gaussian system at $t\rightarrow \infty $. The weaker amplification
range is defined by $\Gamma _3^2+4\eta _1^2<\Gamma _0^2$. }
\end{figure}

\section{The symmetric amplifier}

The system may undergo symmetric single mode amplification as well as the
inter-mode amplification, that is $\eta _3=0,$ $\eta _0\neq 0.$ We consider
the situation of weak amplification, that is, $C_2>B_1$ ($\eta _0>0,$ $%
\Gamma _3>0$ is assumed). When $t\rightarrow \infty $, we have $M\rightarrow
0,N\rightarrow 0$, the final state is specified by the residue matrices $%
\alpha $ and $\beta $ (see Appendix). It seems that the separability
criterion might be very complicate, however, a direct calculation shows that
the condition can be written as a quadrature form of the square of the
inter-mode amplification parameter $\eta _1,$
\begin{equation}
s_2\eta _1^4+s_1\eta _1^2+s_0\geq 0,
\end{equation}
with $s_0=(1-\eta _0^{\prime 2})^2[\eta _0^{\prime 4}+8(1+\Gamma _3^{\prime
2})\eta _0^{\prime 2}\overline{n}_0(\overline{n}_0+1)+16(1-\Gamma _3^{\prime
2})^2\overline{n}_0^2(\overline{n}_0+1)^2],$ $s_2=[1-\eta _0^{\prime
2}-\Gamma _3^{\prime 2}(2\overline{n}_0+1)^2]^2,$ $s_1=-2\eta _0^{\prime
6}-8\eta _0^{\prime 4}\overline{n}_0(\overline{n}_0+1)-2\Gamma _3^{\prime
2}\eta _0^{\prime 4}(8\overline{n}_0^2+8\overline{n}_0+1)$ $-4(1-\Gamma
_3^{\prime 2})(1-\Gamma _3^{\prime 2}(2\overline{n}_0+1)^2)(2\overline{n}%
_0^2+2\overline{n}_0+1)$ $+2\eta _0^{\prime 2}[8\overline{n}_0^2+8\overline{n%
}_0+3-4\Gamma _3^{\prime 4}\overline{n}_0(\overline{n}_0+1)(2\overline{n}%
_0+1)^2$ $+\Gamma _3^{\prime 2}(16\overline{n}_0^4+32\overline{n}_0^3+24%
\overline{n}_0^2+8\overline{n}_0-1)].$

The border of the separable state set and entangled state set is shown in
Fig.2 with $\eta _0=0.5$ , where only the case of weak amplification is
shown. Our numerical result shows that the range (in terms of relative
asymmetric damping quantity $\Gamma _3^{\prime }=\Gamma _3/\Gamma _0=(\Gamma
_1-\Gamma _2)/(\Gamma _1+\Gamma _2)$ and the noise $\overline{n}_0$) of weak
amplification shrinks as $\eta _0$ increasing, the weak amplification
entanglement can only be possible when the noise is less than $1/2$ photon
number.
\begin{figure}[tbp]
\includegraphics[ trim=0.000000in 0.000000in -0.138042in 0.000000in,
height=2.0081in, width=2.5097in ]{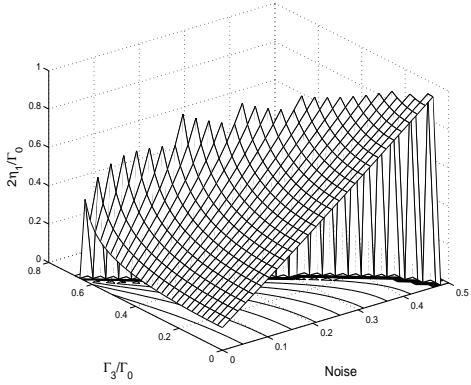}
\caption{ The separability border of the symmetric single-mode
amplification, inter-mode amplification and non-symmetric damping Gaussian
system at $t\rightarrow \infty $. The weaker amplification case.}
\end{figure}

\section{Conclusion}

We have studied the evolution of two-mode Gaussian state under
non-symmetric damping, symmetric amplification and thermal noise.
The non-symmetric damping is the most general damping of two-mode
system. The amplification is limited the symmetric case for
simplicity, although the most general case of $\eta _0\neq 0,\eta
_1\neq 0,$ $\eta _3\neq 0$ is also solvable. The case of
inter-mode amplification alone is especially simple, its
separability criterions of final states in both weak and strong
amplifications were obtained. The separability criterion of the
final state of symmetric amplification ($\eta _0\neq 0,\eta _1\neq
0,$ $\eta _3=0$) is given for weak amplification. Inter-mode
amplification parameter $\eta _1$ is crucial for entanglement.

In the weak amplification case, final state entanglement is only possible
when the thermal noise $\overline{n}_0$ is less than $1/2$ photon number.
When the single mode amplification parameter $\eta _0$ increases, the
entanglement range in terms of relative asymmetric damping quantity $\Gamma
_3^{\prime }=\Gamma _3/\Gamma _0=(\Gamma _1-\Gamma _2)/(\Gamma _1+\Gamma _2)$%
, the noise $\overline{n}_0$ and inter-mode amplification normalized
parameter $\eta _1^{\prime }$ shrinks. Lower $\Gamma _3^{\prime },$ $%
\overline{n}_0$ and higher $\eta _1^{\prime }$ are required for the state to
be entangled when $\eta _0$ increases.

\section*{Appendix: The residue matrices $\alpha $ and $\beta $}

In the two mode situation, denote $\alpha =\sum_{i=0}^3\alpha _i\sigma _i,$
all $\alpha _i$ are real due to $\alpha ^{\dagger }=\alpha ;$ denote $\beta
=\sum_{i=0,1,3}\beta _i\sigma _i,\beta _i=\beta _{iR}+i\beta _{iI},$ the $%
\sigma _2$ item is nullified due to $\beta ^T=\beta .$ Let $\Gamma
_{0,3}=\frac 12(\Gamma _1\pm \Gamma _2),$ then $\Gamma =\Gamma _0\sigma
_0+\Gamma _3\sigma _3.$ Together with $\eta =\sum_{i=0,1,3}\eta _i\sigma _i$
and $\overline{n}=\overline{n}_0\sigma _0$, all the matrices in Eqs.(\ref
{we3}) (\ref{we4}) are expressed in the basis of Pauli matrices. By
comparing the coefficient of the Pauli matrices, from Eqs.(\ref{we3}) (\ref
{we4}), we obtain two groups of equations, the first group equations
containing $\mathbf{\alpha =}(\alpha _0,\alpha _1,\alpha _3)^T,$ $\mathbf{%
\beta }_R=(\beta _{0R},\beta _{1R},\beta _{3R})^T$ are
\begin{eqnarray*}
G\mathbf{\alpha -}E\mathbf{\beta }_R &=&(\overline{n}_0+\frac 12)(\Gamma
_0,0,\Gamma _3)^T \\
E\mathbf{\alpha -}G\mathbf{\beta }_R &=&\mathbf{0}
\end{eqnarray*}
with
\[
G=\left[
\begin{array}{lll}
\Gamma _0 & 0 & \Gamma _3 \\
0 & \Gamma _0 & 0 \\
\Gamma _3 & 0 & \Gamma _0
\end{array}
\right] ,\text{ }E=2\left[
\begin{array}{lll}
\eta _0 & \eta _1 & \eta _3 \\
\eta _1 & \eta _0 & 0 \\
\eta _3 & 0 & \eta _0
\end{array}
\right] .
\]
The second group equations containing $(\alpha _2,\beta _{0I},\beta
_{1I},\beta _{3I})$ have a solution $(\alpha _2,\beta _{0I},\beta
_{1I},\beta _{3I})=$ $\mathbf{0.}$ The solution to Eqs.(\ref{we3}) (\ref{we4}%
) is
\begin{eqnarray*}
\mathbf{\alpha } &=&(\overline{n}_0+\frac 12)(G-EG^{-1}E)^{-1}(\Gamma
_0,0,\Gamma _3)^T \\
\mathbf{\beta } &=&G^{-1}E\mathbf{\alpha }
\end{eqnarray*}
When $\eta _3=0,$ the solution is
\begin{eqnarray*}
\alpha &=&(\overline{n}_0+\frac 12)\Delta ^{-1}\{(\Gamma _0^2-4\eta
_0^2)[(\Gamma _0^2-\Gamma _3^2)^2+4\Gamma _3^2(\eta _1^2-\eta _0^2)-4\Gamma
_0^2(\eta _1^2+\eta _0^2)]\sigma _0 \\
&&+4\eta _0\eta _1[(2\Gamma _0^2-\Gamma _3^2)(\Gamma _0^2-\Gamma
_3^2)+4\Gamma _3^2(\eta _1^2-\eta _0^2)-8\Gamma _0^2\eta _0^2]\sigma _1 \\
&&+\Gamma _0\Gamma _3[16(2\eta _0^2-\eta _1^2)(\eta _0^2-\eta _1^2)+4\eta
_1^2(\Gamma _3^2-\Gamma _0^2)-8\Gamma _0^2\eta _0^2]\sigma _3\},
\end{eqnarray*}
\begin{eqnarray*}
\beta &=&(\overline{n}_0+\frac 12)\Delta ^{-1}\{2\Gamma _0\eta _0(\Gamma
_0^2-4\eta _0^2)[\Gamma _0^2-\Gamma _3^2+4(\eta _1^2-\eta _0^2)]\sigma _0 \\
&&+2\Gamma _0\eta _1[(\Gamma _0^2-\Gamma _3^2)^2+8\eta _0^2(2\eta _1^2-2\eta
_0^2-\Gamma _3^2)+4\eta _1^2(\Gamma _3^2-\Gamma _0^2)]\sigma _1 \\
&&+2\eta _0\Gamma _3[16(\eta _0^2-\eta _1^2)^2+\Gamma _0^2(\Gamma
_3^2-\Gamma _0^2-8\eta _1^2)+4\Gamma _3^2(\eta _1^2-\eta _0^2)]\sigma _3\},
\end{eqnarray*}
where $\Delta =(\Gamma _0^2-4\eta _0^2)[(\Gamma _0^2-\Gamma _3^2-4\eta
_0^2-4\eta _1^2)^2-4\eta _0^2(\Gamma _3^2+4\eta _1^2)]=(\Gamma _0^2-4\eta
_0^2)[(\Gamma _0+2\eta _0)^2-(\Gamma _3^2+4\eta _1^2)][(\Gamma _0-2\eta
_0)^2-(\Gamma _3^2+4\eta _1^2)].$ When $\eta _0=\eta _3=0,$ the solution is
\begin{eqnarray*}
\alpha &=&\frac{(\overline{n}_0+\frac 12)}{\Gamma _0\left( \Gamma
_0^2-\Gamma _3^2-4\eta _1^2\right) }[\Gamma _0(\Gamma _0^2-\Gamma
_3^2)\sigma _0+4\Gamma _3\eta _1^2\sigma _3], \\
\beta &=&\frac{2(\overline{n}_0+\frac 12)\eta _1(\Gamma _0^2-\Gamma _3^2)}{%
\Gamma _0\left( \Gamma _0^2-\Gamma _3^2-4\eta _1^2\right) }\sigma _1.
\end{eqnarray*}

\section*{Acknowledgment}

Funding by the National Natural Science Foundation of China (under Grant No.
10575092), Zhejiang Province Natural Science Foundation (under Grant No.
RC104265) and AQSIQ of China (under Grant No. 2004QK38) are gratefully
acknowledged.

\end{document}